\documentclass[10pt,english,prd,aps,showpacs,epsf,floats]{revtex4}
\usepackage[T1]{fontenc}
\usepackage[latin1]{inputenc}
\usepackage{amsmath}
\usepackage{amssymb}
\usepackage{amsfonts}
\usepackage{babel}

\makeatletter
\makeatletter
\input{tcilatex}
\makeatother
\makeatother

\begin{document}

\title{Information Geometric Modeling of Scattering Induced Quantum
Entanglement}
\author{D.-H. Kim$^{1,2}$, S. A. Ali$^{2,3,4}$, C. Cafaro$^{5}$, S. Mancini$%
^{6}$}
\affiliation{$^{1}$Institute for the Early Universe, Ewha Womans University, Seoul
120-750, South Korea\\
$^{2}$International Institute for Theoretical Physics and Mathematics
Einstein-Galilei, via Santa Gonda 14, 59100 Prato, Italy\\
$^{3}$Department of Physics, State University of New York at Albany, 1400
Washington Avenue, Albany, NY 12222, USA\\
$^{4}$Department of Arts and Sciences, Albany College of Pharmacy and Health
Sciences, 106 New Scotland Avenue, Albany, NY 12208, USA\\
$^{5\text{, }6}$School of Science and Technology, Physics Division,
University of Camerino, I-62032 Camerino, Italy}

\begin{abstract}
We present an information geometric analysis of entanglement generated by an%
\textbf{\ }$s$\textbf{-}wave scattering between two Gaussian wave packets.
We conjecture that the pre and post-collisional quantum dynamical scenarios
related to an elastic head-on collision are macroscopic manifestations
emerging from microscopic statistical structures. We then\textbf{\ }describe
them by uncorrelated and correlated Gaussian statistical models,
respectively. This allows us to express the entanglement strength in terms
of scattering potential and incident particle energies. Furthermore, we show
how the entanglement duration can be related\textbf{\ }to the scattering
potential and incident particle energies. Finally, we discuss the connection
between entanglement and complexity of motion.
\end{abstract}

\pacs{%
Probability
Theory
(02.50.Cw),
Riemannian
Geometry
(02.40.Ky),
Complexity
(89.70.Eg),
Entropy
(89.70.Cf), Quantum
Entanglement
(03.65.Ud).%
}
\maketitle

One of the most important features of composite quantum mechanical systems
is their ability to become entangled \cite{epr, Schrodinger}. In general,
quantum entanglement is described by quantum correlations among the distinct
subsystems of the entire composite quantum system. For such correlated
quantum systems, it is not possible to specify the quantum state of any
subsystem independently of the remaining subsystems \cite{epr}. Apart from
these remarks, the fundamental meaning of quantum entanglement is still a
widely debated issue \cite{brukner}.

From a conceptual point of view, the simplest and most realistic mechanism
of generating entanglement between two particles is via scattering processes 
\cite{law, Bubhardt}. The two particles can become entangled as they
approach each other as a consequence of mutual interactions. For instance,
for interaction potentials with a strong repulsive core, quantum
interference between incident and reflected waves can generate transient
entanglement. After the collision, the two particles may still be entangled
and share forms of quantum information in the final scattered state. Quantum
entanglement can also be generated during inelastic collisions between the
dissipative walls of a container and the quantum system confined within it 
\cite{schulman}. Entanglement may also be induced in multi-atom systems
confined in a harmonic trap interacting via a delta interaction potential 
\cite{mack}.

In order to obtain a clear and detailed understanding of entanglement, it is
first necessary to quantify it. It is known that for pure maximally
entangled states the quantum state of any subsystem is maximally mixed,
while for separable states it is pure. Thus, one is led to consider the von
Neumann entropy of the reduced state, measuring its degree of mixedness, as
an entanglement measure. This turns out to be correct for pure bipartite
states \cite{popescu} (the case we are considering in this Letter), while
for more general states other entanglement measures should be invoked \cite%
{horodecki, plenio}. Apart from the above presented remarks, a great deal
remains unclear about the physical interpretation of entanglement measures 
\cite{wootters} and much remains unsatisfactory about our understanding of
scattering-induced quantum entanglement, especially with regard to how
interaction potentials and particle energies control the entanglement \cite%
{law}. Finally, our knowledge of the connections between entanglement and
complexity of motion remains far from complete \cite{casati}.

In this Letter we investigate the potential utility of the \emph{Information
Geometric Approach to Chaos} (IGAC) \cite{cafarothesis, cafaroCSF} in
characterizing the quantum entanglement produced by a head-on elastic
collision between two Gaussian wave packets interacting via a scattering
process\ \cite{Wang}.

IGAC is a theoretical framework developed to study the complexity of
informational geodesic flows on curved statistical manifolds underlying the
probabilistic description of physical, biological or chemical systems. It is
the information geometric analogue of conventional geometrodynamical
approaches \cite{casetti, di bari} where the classical configuration space
is replaced by a statistical manifold with the additional possibility of
considering complex dynamics arising from non conformally flat metrics (the
Jacobi metric is always conformally flat). For recent applications of the
IGAC\ to quantum physics we refer to \cite{cafaroMPLB, cafaroPA}.

Here we conjecture that the scattering induced quantum entanglement is a
macroscopic manifestation emerging from specific statistical
microstructures. Specifically, using information geometric techniques \cite%
{amari} and inductive inference methods \cite{giffin, caticha}, we propose
that the pre and post-collisional scenarios are modelled by an uncorrelated 
\cite{cafaroPD} and correlated Gaussian statistical model \cite{cafaroPA2010}%
, respectively. We present an analytical connection between the entanglement
strength - quantified in terms of purity - to the scattering potential and
incident particle energies. Furthermore, we relate the entanglement duration
to the scattering potential and incident particle energies. Finally, we
uncover a quantitative relation between quantum entanglement and the
information geometric complexity of motion \cite{cafaroAMC}.

Before describing the physical system being studied, we recall that
spatially localized Gaussian wave packets are especially useful to describe
naturally occurring quantum states and they are easy to handle since many
important properties of these states can often be obtained in an analytic
fashion \cite{klauder}. Furthermore, the Wigner distribution of Gaussian
wave packets is positive-definite and therefore Gaussian states could be
tagged as essentially classical \cite{hillery}.

The physical system being considered consists of two interacting spin-$0$
particles of equal mass $m$. For such a system, a complete set of commuting
observables is furnished by the momentum operators of each particle \cite%
{Harshman}. In terms of the center of mass and relative coordinates, the
Hamiltonian $\mathcal{H}$ of the system becomes \cite{Wang},%
\begin{equation}
\mathcal{H}=\mathcal{H}_{\text{cm}}+\mathcal{H}_{\text{rel}}\text{,}
\end{equation}%
with%
\begin{equation}
\mathcal{H}_{\text{cm}}=\frac{P^{2}}{2M}\text{ \ and, \ }\mathcal{H}_{\text{%
rel}}=\frac{p^{2}}{2\mu }+V\left( x\right) \text{,}  \label{due}
\end{equation}%
where $M\equiv 2m$ is the total mass and $\mu \equiv \frac{m}{2}$ is the
reduced mass. The interaction potential $V\left( x\right) $ is isotropic and
has a short range $d$ such that $V\left( x\right) \approx 0$ for $x>d$.
Before colliding, the two particles are in the form of disentangled Gaussian
wave packets, each characterized by a width $\sigma _{0}$ in momentum space.
The initial distance between the two particles is $R_{0}$ and their average
initial momenta - setting the Planck constant $\hbar $ equal to one - are $%
\mp k_{0}$, respectively. We emphasize that the three-dimensional scattering
process in \cite{Wang} can be effectively reduced to a one-dimensional
process as far as the probability density-based analysis concerns. This can
be explained by observing that both the pre and post-collisional
three-dimensional Gaussian wave-packets in \cite{Wang} are isotropic. That
is to say, in spherical coordinates the representations of the separable
two-particle states exhibit a functional dependence on the radial variable
only. For this reason the three-dimensional vectorial representations may be
effectively reduced to one-dimensional representations. After some algebra,
it follows that the initial (pre-collisional) two-particle square wave
amplitude in momentum space reads,%
\begin{equation}
P_{\text{pre}}^{\text{(QM)}}\left( k_{1}\text{, }k\text{$_{2}|k_{0}$, }-k_{0}%
\text{, }\sigma _{0}\right) =\frac{1}{2\pi \sigma _{0}^{2}}\exp \left[ -%
\frac{\left( k_{1}-k_{0}\right) ^{2}+\left( k_{2}+k_{0}\right) ^{2}}{2\sigma
_{0}^{2}}\right] \text{,}  \label{preqm}
\end{equation}%
where $\pm k_{0}$ are the expected values of the momenta $k_{1}$ and $k_{2}$%
, respectively. The square root of the variance for each momentum is denoted
by $\sigma _{0}$. Similarly, following \cite{Wang} and our above mentioned
observation on the dimensional reduction of the analysis, after some tedious
algebra it turns out that the final (post-collisional) two-particle square
wave amplitude in momentum space in the low energy\textbf{\ }$s$-wave
scattering approximation becomes,%
\begin{equation}
P_{\text{post}}^{\text{(QM)}}\left( k_{1}\text{, }k_{2}|\text{$k_{0}$, }%
-k_{0}\text{, }\sigma _{0}\text{; }r_{\text{QM}}\right) =\frac{\exp \left\{ -%
\frac{1}{2\sigma _{0}^{2}\left( 1-r_{\text{QM}}^{2}\right) }\left[ \left(
k_{1}-\text{$k_{0}$}\right) ^{2}-2r_{\text{QM}}\left( k_{1}-\text{$k_{0}$}%
\right) \left( k_{2}+\text{$k_{0}$}\right) +\left( k_{2}+\text{$k_{0}$}%
\right) ^{2}\right] \right\} }{2\pi \sigma _{0}^{2}\sqrt{1-r_{\text{QM}}^{2}}%
}\text{,}  \label{postqm}
\end{equation}%
where the adimensional coefficient $r_{\text{QM}}$ is defined as,%
\begin{equation}
r_{\text{QM}}=r_{\text{QM}}\left( k_{0}\text{, }\sigma _{0}\text{, }R_{0}%
\text{, }a_{S}\right) \overset{\text{def}}{=}\sqrt{8\left( 2k_{0}^{2}+\sigma
_{0}^{2}\right) R_{0}a_{s}}\text{,}
\end{equation}%
where\textbf{\ }$r_{\text{QM}}\ll 1$\textbf{\ }and $a_{s}$ is the $s$\textbf{%
-}wave scattering length. As a side remark, we point out that (\ref{postqm})
reduces to (\ref{preqm}) when\textbf{\ }$r_{\text{QM}}$ vanishes.

As pointed out earlier, in order to properly analyze entanglement, the
entanglement entropy obtained from the long time limit post-collisional wave
function is required. In most cases however, this must be performed
numerically. Thus, to approach the problem analytically and simultaneously
gain insights into the problem, it is convenient to make use of the
linearized version of the entropy of the system, i.e. of the purity of the
system \cite{Wang}. The purity function is defined as $\mathcal{P}\overset{%
\text{def}}{=}$Tr$\left( \rho _{A}^{2}\right) $ where $\rho _{A}\equiv $ Tr$%
_{B}\left( \rho _{AB}\right) $ is the reduced density matrix of particle $A$
and $\rho _{AB}$ is the two-particle density matrix associated with the
post-collisional two-particle wave function. For pure two-particle states,
the smaller the value of $\mathcal{P}$ the higher the entanglement. That is,
the loss of purity provides an indicator of the degree of entanglement.
Hence, a disentangled product state corresponds to $\mathcal{P}=1$. We
emphasize that the purity has been used as a measure of the degree of
entanglement in various physical situations \cite{jac}, especially in atomic
physics in order to characterize the two-body correlations in dynamical
atomic processes \cite{liu, wagner}. Under the assumption that the two
particles are well separated both initially (before collision) and finally
(after collision) \cite{merz} and assuming that the colliding Gaussian wave
packets are very narrow in the momentum space ($\sigma _{0}\ll 1$ such that
the phase shift can be treated as a constant $\theta \left( k_{0}\right)
\equiv \theta _{0}$), it follows that the purity of the post-collisional
two-particle wave function is approximately given by \cite{Wang}%
\begin{equation}
\mathcal{P}=1-\frac{4\left( 2k_{0}^{2}+\sigma _{_{0}}^{2}\right) R_{0}\sqrt{%
S_{0}}}{\sqrt{\pi }}+\mathcal{O}\left( S_{0}\right) \text{,}  \label{pure}
\end{equation}%
where $S_{0}\overset{\text{def}}{=}4\pi \left\vert f\left( k_{0}\right)
\right\vert ^{2}$ is the scattering cross section and $f\left( k_{0}\right) =%
\frac{e^{i\theta _{0}}\sin \theta _{0}}{k_{0}}\approx \frac{\theta _{0}}{%
k_{0}}$ is the $s$-wave scattering amplitude.

Although very important, the analysis of \cite{Wang} does not address the
problem of how the interaction potentials and particle energies control the
scattering-induced entanglement (a key problem as pointed out in \cite{law})
and it does not discuss any possible connection between the entanglement
generated in the scattering process and the complexity of the motion related
to the pre and post-collisional quantum dynamical scenarios (a key problem
as pointed out in \cite{casati}).

In this Letter, we attempt to provide some answers to such unsolved relevant
issues. We conjecture that the pre and post-collisional quantum dynamical
scenarios characterized by (\ref{preqm}) and (\ref{postqm}), respectively,
and describing the quantum entanglement (quantified in terms of the purity $%
\mathcal{P}$ in (\ref{pure})) produced by a head-on collision between two
Gaussian wave packets are macroscopic manifestations emerging from specific
underlying microscopic statistical structures. We stress that within the
IGAC, we provide a probabilist description of physical systems by studying
the temporal evolution on curved statistical manifolds of probability
distributions encoding all the relevant available information concerning the
system considered. For the problem under investigation, we propose that $P_{%
\text{pre}}^{\text{(QM)}}\left( k_{1}\text{, }k\text{$_{2}|k_{0}$, }-k_{0}%
\text{, }\sigma _{0}\right) $ can be interpreted as a limiting case (initial
time limit) arising from an evolving uncorrelated Gaussian probability
distribution $P_{\text{pre}}^{\text{(IG)}}\left( k_{1}\text{, }k_{2}|\mu
_{k_{1}}\left( \tau \right) \text{, }\mu _{k_{2}}\left( \tau \right) \text{, 
}\sigma \left( \tau \right) \right) $,%
\begin{equation}
P_{\text{pre}}^{\text{(IG)}}\left( k_{1}\text{, }k_{2}|\mu _{k_{1}}\left(
\tau \right) \text{, }\mu _{k_{2}}\left( \tau \right) \text{, }\sigma \left(
\tau \right) \right) \overset{\text{def}}{=}\frac{\exp \left\{ -\frac{1}{%
2\sigma ^{2}}\left[ \left( k_{1}-\mu _{k_{1}}\right) ^{2}+\left( k_{2}-\mu
_{k_{2}}\right) ^{2}\right] \right\} }{2\pi \sigma ^{2}}\text{.}
\label{igpre}
\end{equation}%
As a matter of fact, setting $\mu _{k_{1}}\left( 0\right) =k_{0}$, $\mu
_{k_{2}}\left( 0\right) =-k_{0}$ and $\sigma \left( 0\right) =\sigma _{0}$,
we obtain%
\begin{equation}
P_{\text{pre}}^{\text{(IG)}}\left( k_{1}\text{, }k\text{$_{2}|k_{0}$, }-k_{0}%
\text{, }\sigma _{0}\right) =P_{\text{pre}}^{\text{(QM)}}\left( k_{1}\text{, 
}k\text{$_{2}|k_{0}$, }-k_{0}\text{, }\sigma _{0}\right) \text{.}  \label{A}
\end{equation}%
Similarly, we propose that $P_{\text{post}}^{\text{(QM)}}\left( k_{1}\text{, 
}k_{2}|\text{$k_{0}$, }-k_{0}\text{, }\sigma _{0}\text{; }r_{\text{QM}%
}\right) $ can be viewed as a limiting case (final time limit) arising from
a correlated Gaussian probability distribution $P_{\text{post}}^{\text{(IG)}%
}\left( k_{1}\text{, }k_{2}|\mu _{k_{1}}\left( \tau \right) \text{, }\mu
_{k_{2}}\left( \tau \right) \text{, }\sigma \left( \tau \right) \text{; }r_{%
\text{IG}}\right) $,%
\begin{equation}
P_{\text{post}}^{\text{(IG)}}\left( k_{1}\text{, }k_{2}|\mu _{k_{1}}\left(
\tau \right) \text{, }\mu _{k_{2}}\left( \tau \right) \text{, }\sigma \left(
\tau \right) \text{; }r_{\text{IG}}\right) =\frac{\exp \left\{ -\frac{\left[
\left( k_{1}-\mu _{k_{1}}\right) ^{2}-2r_{\text{IG}}\left( k_{1}-\mu
_{k_{1}}\right) \left( k_{2}-\mu _{k_{2}}\right) +\left( k_{2}-\mu
_{k_{2}}\right) ^{2}\right] }{2\sigma ^{2}\left( 1-r_{\text{IG}}^{2}\right) }%
\right\} }{2\pi \sigma ^{2}\sqrt{1-r_{\text{IG}}^{2}}}\text{,}
\label{igpost}
\end{equation}%
where $r_{\text{IG}}\overset{\text{def}}{=}\frac{\left\langle
k_{1}k_{2}\right\rangle -\left\langle k_{1}\right\rangle \left\langle
k_{2}\right\rangle }{\sigma ^{2}}$ is the correlation coefficient. Indeed,
setting $\mu _{k_{1}}\left( \tau _{\text{final}}\right) =k_{0}$, $\mu
_{k_{2}}\left( \tau _{\text{final}}\right) =-k_{0}$ and $\sigma \left( \tau
_{\text{final}}\right) =\sigma _{0}$, we obtain 
\begin{equation}
P_{\text{post}}^{\text{(IG)}}\left( k_{1}\text{, }k_{2}|\text{$k_{0}$, }%
-k_{0}\text{, }\sigma _{0}\text{; }r_{\text{IG}}\right) =P_{\text{post}}^{%
\text{(QM)}}\left( k_{1}\text{, }k_{2}|\text{$k_{0}$, }-k_{0}\text{, }\sigma
_{0}\text{; }r_{\text{QM}}\right)  \label{B}
\end{equation}%
in the limit of weak correlations with\textbf{\ }$r_{\text{IG}}\equiv r_{%
\text{QM}}\ll 1$.

At this stage our conjecture is only mathematically sustained by the formal
identities (\ref{A}) and (\ref{B}). To render our conjecture also physically
motivated, recall that $s$-wave scattering can also be described in terms of
a scattering potential $V(x)$ and the scattering phase shift $\theta \left(
k\right) $. For the problem under consideration, $V(x)$ equals the constant
value $V$ for $0\leq x\leq d$ while it is vanishing for $x>d$. The
quantities $V$ and $d$ denote the height (for $V>0$; repulsive potential) or
depth (for $V<0$; attractive potential) and range of the potential,
respectively. Within the standard quantum framework, integrating the radial
part of Schrödinger equation with this potential for the scattered wave and
imposing the matching condition at $x=d$ for its solution and its first
derivative leads to \cite{krane}%
\begin{equation}
\tan \theta _{0}=\frac{k_{\text{out}}\tan \left( k_{\text{in}}d\right) -k_{%
\text{in}}\tan \left( k_{\text{out}}d\right) }{k_{\text{in}}+k_{\text{out}%
}\tan \left( k_{\text{out}}d\right) \tan \left( k_{\text{in}}d\right) }\text{%
,}  \label{match}
\end{equation}%
with $k_{\text{in}}=\sqrt{2\mu \left( T-V\right) }$ for$\;0\leq x\leq d$, $%
k_{\text{out}}=\sqrt{2\mu T}$ for $x>d$ and $\theta _{0}\approx -k_{0}a_{s}$
denotes the $s$\textbf{-}wave scattering phase shift. The quantities $\mu $
and $T$ are the reduced mass and kinetic energy of the two-particle system
in the relative coordinates, respectively; $k_{\text{in}}$ and $k_{\text{out}%
}$ represent\textbf{\ }the conjugate-coordinate wave vectors inside and
outside the potential region, respectively. Equation (\ref{match}) indicates
that the scattering potential $V(x)$ shifts the phase of the scattered wave
at points beyond the scattering region.

Within the IGAC, given the curved statistical manifolds of probability
distributions in (\ref{igpre})\ and (\ref{igpost}), we compute the
Fisher-Rao information metrics on each manifold. We then integrate the two
sets of geodesic equations (three equations for each set) leading to the
expected trajectories connecting the initial and final macroscopic
configurations defined by the macroscopic variables $\left( \mu
_{k_{1}}\left( \tau \right) \text{, }\mu _{k_{2}}\left( \tau \right) \text{, 
}\sigma \left( \tau \right) \right) $\textbf{. }In particular, we obtain that%
\begin{equation}
k_{\text{in}}=\sqrt{1-r_{\text{IG}}}k_{\text{out}}\text{.}  \label{K}
\end{equation}%
Considering Eq. (\ref{K}) together with the limit of low energy $s$-wave
scattering $\left( k_{0}d\ll 1\right) $ and low correlations ($r_{\text{IG}%
}\ll 1$), the matching condition (\ref{match}) in the information geometric
framework reads%
\begin{equation}
\theta \left( k_{0}\right) \approx -\frac{1}{3}r_{\text{IG}}d^{3}k_{0}^{3}%
\text{,}  \label{tetta}
\end{equation}%
where%
\begin{equation}
r_{\text{IG}}=\frac{V}{T}=\frac{2\mu V}{k_{0}^{2}}\text{.}  \label{rIG}
\end{equation}%
Combining (\ref{tetta}) and (\ref{rIG}), we obtain%
\begin{equation}
\theta \left( k_{0}\right) \approx -\frac{2}{3}\mu Vd^{3}k_{0}\text{.}
\label{tettona}
\end{equation}%
Equation (\ref{tettona}), obtained via information geometric dynamical
methods, is in perfect agreement with the result presented in \cite{Mishima}
(and not in \cite{Wang}) where standard Schrodinger's quantum dynamics was
employed. Such finding is especially important because it allows to state
that our conjecture is also physically motivated.

As a consequence of (\ref{pure}) and (\ref{tetta}), we find that when both
low energy and weak correlation regimes occur, the purity $\mathcal{P}$ of
the system becomes%
\begin{equation}
\mathcal{P}\approx 1-\frac{16\mu V\left( 2k_{0}^{2}+\sigma
_{_{0}}^{2}\right) R_{0}d^{3}}{3}\text{. }  \label{p29}
\end{equation}%
Equation (\ref{p29}) implies that the purity $\mathcal{P}$ can be expressed
in terms of physical quantities such as the scattering potential $V\left(
x\right) $ and the initial quantities $k_{0}$, $\sigma _{0}$ and $R_{0}$ via
(\ref{rIG}). Apart from (\ref{tettona}), Eq. (\ref{p29}) is the first
significant finding obtained within our hybrid approach (quantum dynamical
results combined with information geometric modeling techniques) that allows
to explain how the interaction potential $V\left( x\right) $ and the
incident particle energies $T$ control the strength of the entanglement. The
role played by $r_{\text{IG}}$ in the quantities $\mathcal{P}$ and $V$
suggests that the physical information about quantum scattering and
therefore about quantum entanglement is encoded in the statistical
correlation coefficient, specifically in the covariance term $\mathrm{Cov}%
\left( k_{1}\text{, }k_{2}\right) \overset{\text{def}}{=}\left\langle
k_{1}k_{2}\right\rangle -\left\langle k_{1}\right\rangle \left\langle
k_{2}\right\rangle $ appearing in the definition of $r_{\text{IG}}$.

Within the IGAC we are also able to estimate the statistical temporal
duration over which the entanglement is active. Indeed, it turns out that
the uncorrelated and the correlated statistical Gaussian models would
require time intervals $\tau _{\text{uncorr.}}$ and $\tau _{\text{corr.}}$,
respectively, to attain the same value as the initial momentum\textbf{\ }$%
k_{0}$. Assuming $r_{\text{IG}}\ll 1$, from the above-mentioned geodesic
equation analysis it can be shown that what we define "\emph{entanglement
duration}" $\Delta \left( k_{0}\text{, }\sigma _{0}\text{, }r_{\text{IG}}%
\text{ }\right) $ reads, 
\begin{equation}
\Delta \left( k_{0}\text{, }\sigma _{0}\text{, }r_{\text{IG}}\text{ }\right) 
\overset{\text{def}}{=}\tau _{\text{corr.}}-\tau _{\text{uncorr.}}\propto
\left\vert \ln \left\{ 1-\left[ \left( 1-r_{\text{IG}}\right) ^{-1/2}-1%
\right] \cdot \eta _{\Delta }\left( k_{0}\text{, }\sigma _{0}\right)
\right\} \right\vert \text{,}  \label{duration}
\end{equation}%
where $\eta _{\Delta }=\eta _{\Delta }\left( k_{0}\text{, }\sigma
_{0}\right) $ is given by 
\begin{equation}
\eta _{\Delta }\left( k_{0}\text{, }\sigma _{0}\right) =\left( \frac{k_{0}}{%
\sigma _{0}}\right) ^{2}\exp \left[ \left( \frac{\sigma _{0}}{k_{0}}\right)
^{2}-\frac{3}{4}\left( \frac{\sigma _{0}}{k_{0}}\right) ^{4}+\mathcal{O}%
\left[ \left( \frac{\sigma _{0}}{k_{0}}\right) ^{6}\right] \right] \text{ \
for \ }\frac{\sigma _{0}}{k_{0}}\ll 1\text{.}  \label{eta}
\end{equation}%
Here, we can find the upper bound value of $r_{\text{IG}}$ by means of (\ref%
{duration}) and (\ref{eta}), 
\begin{equation}
r_{\text{IG}}<\frac{2}{\eta _{\Delta }\left( k_{0}\text{, }\sigma
_{0}\right) }\text{.}  \label{r_bound}
\end{equation}%
For example, with $\sigma _{0}/k_{0}\sim 10^{-3}$ we have $r_{\text{IG}%
}<2\times 10^{-6}$. We observe that the entanglement duration can be
controlled via the initial parameters $k_{0}$, $\sigma _{0}$ and the
correlations $r_{\text{IG}}$ (therefore via the incident particle energies
and the scattering potential due to (\ref{rIG})). Also, we notice that in
the absence of correlations, i.e. $r_{\text{IG}}\rightarrow 0$, $\Delta
\rightarrow 0$. It is anticipated that the maximum duration would be
obtained when $r_{\text{IG}}$ is the greatest and the ratio $\sigma
_{0}/k_{0}$ is the smallest. In summary, the entanglement duration allows to
quantitatively estimate the temporal interval over which the entanglement is
active in terms of statistical geodesic paths (statistical evolution of
probability distributions) on the statistical manifolds underlying the
information dynamics used to describe the pre and post-collisional scenarios
(absence and presence of entanglement, respectively). It encodes information
about how long it would take for an entangled system to overcome the
momentum gap (relative to a corresponding non-entangled system) generated by
the scattering phase shift. The entangled system only attains the full value
of momentum (i.e. the momentum value as seen in the corresponding
non-entangled system) when the scattering phase shift vanishes. For this
reason, the entanglement duration represents the statistical temporal
duration over which the entanglement is active.

Our final finding uncovers an interesting quantitative connection between
quantum entanglement quantified by the purity $\mathcal{P}$ in (\ref{p29})
and the information geometric complexity of motion on the uncorrelated and
correlated curved statistical manifolds $\mathcal{M}_{s}^{\left( \text{%
uncorr.}\right) }$ and $\mathcal{M}_{s}^{\left( \text{corr.}\right) }$,
respectively. The information geometric complexity (IGC) represents the
volume of the effective parametric space explored by the system in its
evolution between the chosen initial and final macrostates. In general, the
volume itself is in general given in terms of a multidimensional
fold-integral over the geodesic paths connecting the initial and final
macrostates \cite{cafaroAMC},%
\begin{equation}
\mathcal{C}_{\text{IGC}}\overset{\text{def}}{=}\frac{1}{\tau }\int_{0}^{\tau
}d\tau ^{\prime }\emph{vol}\left[ \mathcal{D}_{\Theta }^{\left( \text{%
geodesic}\right) }\left( \tau ^{\prime }\right) \right] \text{,}  \label{rhs}
\end{equation}%
where%
\begin{equation}
\emph{vol}\left[ \mathcal{D}_{\Theta }^{\left( \text{geodesic}\right)
}\left( \tau ^{\prime }\right) \right] \overset{\text{def}}{=}\int_{\mathcal{%
D}_{\Theta }^{\left( \text{geodesic}\right) }\left( \tau ^{\prime }\right)
}\rho _{\left( \mathcal{M}_{s}\text{, }g\right) }\left( \theta ^{1}\text{%
,..., }\theta ^{n}\right) d^{n}\Theta \text{.}  \label{v}
\end{equation}%
The quantity $\rho _{\left( \mathcal{M}_{s}\text{, }g\right) }\left( \theta
^{1}\text{,..., }\theta ^{n}\right) $ is the so-called Fisher density and
equals the square root of the determinant of the metric tensor $g_{\mu \nu
}\left( \Theta \right) $,%
\begin{equation}
\rho _{\left( \mathcal{M}_{s}\text{, }g\right) }\left( \theta ^{1}\text{%
,..., }\theta ^{n}\right) \overset{\text{def}}{=}\sqrt{g\left( \left( \theta
^{1}\text{,..., }\theta ^{n}\right) \right) }\text{.}
\end{equation}%
The integration space $\mathcal{D}_{\Theta }^{\left( \text{geodesic}\right)
}\left( \tau ^{\prime }\right) $ in (\ref{v}) is defined as follows,%
\begin{equation}
\mathcal{D}_{\Theta }^{\left( \text{geodesic}\right) }\left( \tau ^{\prime
}\right) \overset{\text{def}}{=}\left\{ \Theta \equiv \left( \theta ^{1}%
\text{,..., }\theta ^{n}\right) :\theta ^{k}\left( 0\right) \leq \theta
^{k}\leq \theta ^{k}\left( \tau ^{\prime }\right) \right\} \text{,}
\label{is}
\end{equation}%
where $k=1$,.., $n$ and $\theta ^{k}\equiv \theta ^{k}\left( s\right) $ with 
$0\leq s\leq \tau ^{\prime }$ such that,%
\begin{equation}
\frac{d^{2}\theta ^{k}\left( s\right) }{ds^{2}}+\Gamma _{lm}^{k}\frac{%
d\theta ^{l}}{ds}\frac{d\theta ^{m}}{ds}=0\text{.}
\end{equation}%
The integration space $\mathcal{D}_{\Theta }^{\left( \text{geodesic}\right)
}\left( \tau ^{\prime }\right) $ in (\ref{is}) is a $n$-dimensional subspace
of the whole (permitted) parameter space $\mathcal{D}_{\Theta }^{\left( 
\text{tot}\right) }$. The elements of $\mathcal{D}_{\Theta }^{\left( \text{%
geodesic}\right) }\left( \tau ^{\prime }\right) $ are the $n$-dimensional
macrovariables $\left\{ \Theta \right\} $ whose components $\theta ^{k}$ are
bounded by specified limits of integration $\theta ^{k}\left( 0\right) $ and 
$\theta ^{k}\left( \tau ^{\prime }\right) $ with $k=1$,.., $n$. The limits
of integration are obtained via integration of the $n$-dimensional set of
coupled nonlinear second order ordinary differential equations
characterizing the geodesic equations. In our case $n=3$ with $\Theta \equiv
\left( \mu _{k_{1}}\left( \tau \right) \text{, }\mu _{k_{2}}\left( \tau
\right) \text{, }\sigma \left( \tau \right) \right) $ and following the line
of reasoning presented in \cite{cafaroPD, cafaroPA2010} one finds that%
\begin{equation}
\mathcal{C}_{\text{IGC}}^{\left( \text{corr.}\right) }=\sqrt{\frac{1-r_{%
\text{IG}}}{1+r_{\text{IG}}}}\mathcal{C}_{\text{IGC}}^{\left( \text{uncorr.}%
\right) }\text{,}  \label{complexity}
\end{equation}%
where $\mathcal{C}_{\text{IGC}}^{\left( \text{corr.}\right) }$ and $\mathcal{%
C}_{\text{IGC}}^{\left( \text{uncorr.}\right) }$ denotes the information
geometric complexities of motion on the chosen statistical manifolds. As a
side remark, we point out that (\ref{complexity}) confirms that an increase
in the correlational structure of the dynamical equations for the
statistical variables labelling a macrostate of a system implies a reduction
in the complexity of the geodesic paths on the underlying curved statistical
manifolds \cite{carloPS, carloPD}. In other words, making macroscopic
predictions in the presence of correlations is easier than in their absence.
Combining (\ref{p29}) and (\ref{complexity})\textbf{\ }it follows that%
\begin{equation}
\mathcal{P}\approx 1-\eta _{\mathcal{C}}\left( k_{0}\text{, }\sigma
_{0}\right) \cdot \frac{\Delta \mathcal{C}^{2}}{\mathcal{C}_{\text{total}%
}^{2}}\text{,}  \label{ultima}
\end{equation}%
where,%
\begin{equation}
\Delta \mathcal{C}^{2}\overset{\text{def}}{=}\left[ \mathcal{C}_{\text{IGC}%
}^{\left( \text{uncorr.}\right) }\right] ^{2}-\left[ \mathcal{C}_{\text{IGC}%
}^{\left( \text{corr.}\right) }\right] ^{2}\text{, }\mathcal{C}_{\text{total}%
}^{2}\overset{\text{def}}{=}\left[ \mathcal{C}_{\text{IGC}}^{\left( \text{%
uncorr.}\right) }\right] ^{2}+\left[ \mathcal{C}_{\text{IGC}}^{\left( \text{%
corr.}\right) }\right] ^{2}
\end{equation}%
and, 
\begin{equation}
\eta _{\mathcal{C}}\left( k_{0}\text{, }\sigma _{0}\right) =\frac{8}{3}%
k_{0}^{2}\left( 2k_{0}^{2}+\sigma _{0}^{2}\right) R_{0}d^{3}\text{.}
\end{equation}%
From (\ref{ultima}) it is evident that the scattering-induced quantum
entanglement and the information geometric complexity of motion are
connected. It found that when purity goes to unity, the difference between
the correlated and non-correlated information geometric complexities
approaches zero. In particular, our analysis allows to conclude that the
information geometric complexity of motion (viewed as an indicator of how
"difficult" is to make macroscopic predictions) of Gaussian wave-packets
decreases when the quantum wave-packets becomes entangled. This is
reminiscent of the fact that quantum entanglement is generally considered a
very useful resource at our disposal in quantum computing. As a side remark,
we would like to emphasize that the quantification of the complexity of
quantum motion represents a quite delicate task and it is also a very
debatable issue. Indeed, there is no unique manner in which such complexity
can be described. In particular, the notion of information geometric
complexity employed in this Letter differs from that introduced in \cite%
{casati} where the number of harmonics of the Wigner function was chosen as
a suitable quantum signature of complexity of motion.

In conclusion, our information geometric characterization gives a novel
probabilistic picture of quantum entanglement. Within our framework, the
key-feature is the study of the statistical evolution of classical
probability distributions on curved statistical manifolds underlying the
information dynamics used to describe the pre and post-collisional scenarios
(absence and presence of entanglement, respectively). The emergence of
entanglement manifests itself with a change in the probability distribution
that quantifies our state of knowledge about the system. Once this is
accepted (indeed we provide both mathematical and physical supports to our
conjecture, see Eqs. (\ref{A}), (\ref{B}) and (\ref{tetta}), respectively),
our analysis becomes relevant for the following reasons: first, we are able
to quantitatively express the entanglement strength, quantified by purity,
in terms of scattering potential and incident particle energies (see Eq. (%
\ref{p29})). The scattering potential and incident particle energies are in
turn related to the micro-correlation coefficient $r_{\text{IG}}$, a
quantity that parameterizes the correlated microscopic degrees of freedom of
the system (see Eq. (\ref{rIG})). Second, we introduce a new quantity termed
"entanglement duration" which characterizes the temporal duration over which
the entanglement is active and show that it can be controlled by the initial
momentum $p_{0}$, momentum spread $\sigma _{0}$ and $r_{\text{IG}}$ (see Eq.
(\ref{duration})). Finally, we uncover a quantitative relation between
quantum entanglement and information geometric complexity (see Eq. (\ref%
{ultima})). We also point out that our analysis allows us to interpret
quantum entanglement as a perturbation of statistical space geometry in
analogy to the interpretation of gravitation as perturbation of flat
spacetime. The nature of the perturbation of statistical geometry is global.
This, together with the time-independence of the geometry, leads to the
notion of non-locality. The perturbation of statistical geometry is
associated with the scattering phase shift in the momentum space.

We are confident that the present work represents significant progress
toward the goal of understanding the relationship between statistical
microcorrelations and quantum entanglement on the one hand and the effect of
microcorrelations on the dynamical complexity of informational geodesic
flows on the other. It is our hope to build upon the techniques employed in
this work to ultimately establish a sound information-geometric
interpretation of quantum entanglement together with its connection to
complexity of motion in more general physical scenarios.

\begin{acknowledgments}
This work was supported by WCU (World Class University) program of NRF/MEST
(R32-2009-000-10130-0) and by the European Community's Seventh Framework
Program FP7/2007-2013 under grant agreement 213681 (CORNER Project).
\end{acknowledgments}

\end{document}